# Spatial heterogeneity in earthquake fault-like systems

**J. Kazemian[1], R. Dominguez[2], K.F. Tiampo[1], W. Klein[3]**

[1]{Department of Earth Sciences, Western University, London, Ontario N6A 5B7, Canada}

[2]{Department of Physics, Randolph-Macon College, Ashland, Virginia 23005, USA}

[3]{Department of Physics and Center for Computational Sciences, Boston University, Boston, Massachusetts 02215, USA}

Correspondence to: J. Kazemian (jkazemia@uwo.ca)

**Abstract**

The inhomogeneity of the materials with different physical properties in the Earth is responsible for a wide variety of spatial and temporal behaviors. In this work, we study an earthquake fault model based on Olami-Feder-Christensen (OFC) and Rundle-Jackson-Brown (RJB) cellular automata models with particular aspects of spatial heterogeneities and long-range stress interactions. In our model some localized stress accumulators are added into the system by converting a percentage of randomly selected sites into stronger sites which are called 'asperity cells'. These asperity cells support much higher failure stresses than the surrounding regular lattice sites but eventually rupture when the applied stress reaches their threshold stress. We find that changing the spatial configuration of those stronger sites generally increases capability of the fault system to generate larger events, but that the total percentage of asperities is important as well. We also observe an increasing number of larger events associated with the total number of asperities in the lattice.

# 1   Introduction

Despite the multitude of space-time patterns of activity observed in natural earthquake fault systems, the bulk of the research associated with these patterns has focused on a relatively small fraction of the events, those associated with either larger magnitudes or persistent, localized signals such as aftershock sequences [Kanamori, 1981; Ogata, 1983; Utsu et al., 1995]. One significant problem associated with studies of the earthquake fault network is that the underlying dynamics of the system are not observable [Herz and Hopfield, 1995; Rundle et al., 2000]. A second is that the nonlinear earthquake dynamics are strongly coupled across a wide range of spatial and temporal scales [Kanamori, 1981; Main, 1996; Turcotte, 1997; Rundle et al., 1999; Scholz, 2002]. Finally, the relatively small number of extreme events occur very rarely, impacting our ability to evaluate the significance of the associated local and regional patterns in the instrumental and historic data [Schorlemmer and Gerstenberger, 2007; Vere-Jones, 1995, 2006; Zechar et al., 2010]. As a result, computational simulations are critical to enhancing our understanding of the dynamics of the earthquake system and the occurrence of its largest events [see, e.g., Rundle et al., 2003].

Although simple models cannot replicate the complete spectrum of earthquake phenomenology, these models can provide insights into the important patterns and features associated with the earthquake process and improve our understanding of the dynamics and underlying physics of earthquake fault networks. As a result, simple models of statistical fracture have been used to test some of the typical assumptions and parameters and their possible outcomes (Burridge and Knopoff, 1967; Otsuka, 1972; Rundle and Jackson, 1977; Rundle, 1988; Carlson and Langer, 1989; Nakanishi, 1990, Rundle and Brown, 1991; Olami et al., 1992; Alava et al. 2006). Most of these models assumed a spatially homogeneous earthquake fault, despite the fact that numerical and experimental models of rock fracture suggest that spatial inhomogeneities play an important role in the occurrence of large events and the associated spatial and temporal phenomenology [Dahmen et al., 1998; Turcotte et al., 2003; Tiampo et al., 2002, 2007; Lyakhovsky and Ben-Zion, 2009]. One argument for this approach has been that earthquake faults have long-range stress transfer [Klein et al., 2007]. For long-range stress transfer without inhomogeneities, or randomly distributed inhomogeneities, these models have been found to produce scaling similar to the Gutenberg-Richter (GR) scaling found in real earthquake systems [Gutenberg and Richter, 1956; Serino et al., 2011]. When the stress transfer range is longer than the length scales associated with the inhomogeneities in the system, the dynamics appear to be unaffected by the inhomogeneities. However, recent work by Dominguez et al., [2013] shows that the ratio of the stress transfer range to the length scale of the inhomogeneities affects the GR scaling distribution and the ability of the system to produce large events.

The spatial arrangement of fault inhomogeneities is dependent on the geologic history of the fault. This history is typically quite complex and, as a result, the spatial distribution of the various inhomogeneities occurs on many length scales. Because spatial inhomogeneity plays an important role in the seismicity of an earthquake fault, here we extend the homogeneous OFC model with long-range stress transfer to inhomogeneous models, where particular parameters might vary from site to site.

There have been some earlier studies of the inhomogeneous OFC model (Janosi and Kertesz, 1993; Torvund and Froyland, 1995; Ceva, 1995; Mousseau, 1996; Ramos et al. 2006; Bach et al., 2008; Jagla, 2010). Although, these models considered a number of different ways to impose inhomogeneity on the system, most only investigated systems with nearest neighbor or short-range stress transfer. For example, Janosi and Kertesz (1993) introduced spatial inhomogeneity into the lattice by imposing random site-dependent stress thresholds. Torvund and Froyland (1995) imposed inhomogeneity by changing the uniform distribution of threshold stresses to a Gaussian distribution. Ceva (1995) introduced defects associated with the stress transmission parameter. Ramos (2006) and Jagla (2010) considered varying levels of randomness in the stress threshold.

Serino et al. (2011) studied OFC models with long-range stress transfer in which random damage was incorporated into the lattice. Dominguez et al. (2013) studied the various spatial configurations of damage in a long-range stress transfer model with varying amounts of stress dissipation, $\alpha$ ($0<\alpha\leq 1$) which describe the portion of stress dissipated from the failed sites. Both models represented damage by imposing live and dead sites in the lattice framework.

The live sites can hold an internal stress that is a function of time. Dead sites cannot hold any stress and therefore all the stress that is passed to them during an event is dissipated from the system. The amount of stress dissipated by damage sites at each location can be characterized by the percentage of dead sites in a given neighborhood, $\varphi_i$. Serino et al. (2011) established a connection between the two types of dissipation, stress dissipation and damage dissipation, and Dominguez et al (2013) showed that they can be characterized together in one parameter, $\gamma_i$, herein called site dissipation, $\gamma_i = 1 − \varphi_i(1 − \alpha_i)$. Results showed that both stress dissipation and damage dissipation reduces the length of the scaling regime in their magnitude-frequency distributions and reduces the size of the largest events.

Dominguez et al. (2013) imposed various spatial patterns for the dead sites and compared the behavior with simpler systems with uniformly distributed stress dissipation. Figure 1 shows four different configurations of 25% dead sites for a lattice with linear size L = 256, stress dissipation $\alpha$ = 0, and a stress transfer range of R = 16. These include one case with randomly distributed dead sites (random); a second with blocks of various sizes, where each block has varying values of randomly distributed dead sites (random cascading blocks); a third with completely dead blocks of various sizes (cascading dead blocks); and a fourth with dead blocks of a uniform size (dead blocks). The resulting numerical distribution of events of size s for the various spatial distributions of dead sites shown is shown in Figure 1b. Here we see that the addition of spatial heterogeneity into the lattice *increases* the length of the scaling regime and the size of the largest events as the randomness of the spatial pattern decreases.

Figure 2 investigates the effect of the size of the damage blocks. Again, results are for a lattice with 25% dead sites, a linear size L = 256, stress dissipation $\alpha$ = 0, and a stress transfer range of R = 16. The dead block size, *b*, is varied while the interaction length, R, remains constant. In this case, the scaling regime and maximum event size increases as the ratio R/*b* decreases. The spatial distribution of $\gamma_i$ affects the potential for a large event because failing sites with low values of $\gamma_i$ pass along a high percentage of excess stress to neighboring sites, encouraging additional failures. In the damage only system, site dissipation is determined by spatial locality only, so we require large clumps of sites with low values of $\gamma_i$ in order to allow for a large event.

In this work, the focus is also on the spatial heterogeneity in OFC models with long-range stress transfer. The model is a cellular automata version of earthquake faults based on the OFC (Olami et al., 1992) and RJB (Rundle and Jackson, 1977; Rundle and Brown, 1991) models with some minor variations. Here, inhomogeneities are imposed in the model by allowing a percentage of randomly selected locations that accumulate higher levels of stress, similar to asperities on natural faults. These sites are incorporated by varying the ability of individual sites to support much higher stress for different spatial configurations. We find that the scaling relationship for the heterogeneous systems depends on the amount of the asperity sites as well as the spatial distribution of those asperities and the ratio of the size of the asperities to the stress transfer range. Investigation of the effects of a variety of spatial configurations for asperity sites provides insights into the construction of practical models of an earthquake fault system which is consistent with GR scaling.

## 2  The Model

Our model is a two-dimensional cellular automaton model with periodic boundary conditions. In this model every site in the lattice is connected to $z$ neighbors, which are defined as sites within a certain distance or stress interaction range, $R$. A homogeneous residual stress $\sigma^r$ is assigned to all the sites in the lattice. To impose spatial inhomogeneity on the lattice, two sets of failure thresholds are introduced; 'regular sites' with a constant failure threshold of $\sigma^F$ and 'asperity sites' with a much higher failure threshold ($\sigma^F_{(asperity)} = \sigma^F + \Delta\sigma^F$). These asperity sites are imposed in order to incorporate some percentage of stronger sites into the lattice which will bear higher stress before failure.

Initially, the internal stress variable, $\sigma_j(t)$, is randomly distributed on each site in such a way that the stress on all sites lies between the residual and failure stress thresholds ($\sigma^r < \sigma_i(t=0) < \sigma^F$). At $t=0$ no sites will have $\sigma_i > \sigma^F$. There are several ways to simulate the increase in stress associated with the dynamics of plate tectonics. Here we use the so-called zero velocity limit (Olami et al., 1992). The entire lattice is searched for the site that minimizes ($\sigma^F - \sigma_i$) and that amount of stress is added to each site such that the stress on at least one site is now equal to its failure threshold. That site fails and some fraction of its stress, given by $\alpha [\sigma^F - (\sigma^r \pm \eta)]$, is dissipated from the system. $\alpha$ is a dissipation parameter ($0 < \alpha \leq 1$) which describes the portion of stress dissipated from the failed site and $\eta$ is randomly distributed noise. The failed site's stress is lowered to ($\sigma^r \pm \eta$) and the remaining stress is distributed to its neighbors. After the first site failure, all neighbors are searched to determine if the stress change from the failed site caused any of others to reach their failure stress. If so, the described procedure repeats for those neighbors and if not, the time step (known as the plate update) increases by unity and the lattice is searched again for the next site which minimizes ($\sigma^F - \sigma_i$). The size of the event is calculated from the total number of failures that expand from the first failed site. Stress is dissipated from the system both at the regular lattice sites and through asperity sites which are placed inhomogeneously throughout the lattice. However, because the asperity sites release much higher stress at the time of their failure, the amount of stress dissipated by these sites is different from the bulk of the system and results in inhomogeneous stress dissipation in this model.

Initial results for two different sized large asperity blocks are shown in Figure 3 (a-5% and b-10% of the total lattice sites are considered as asperity sites), with their associated magnitude-frequency relation plotted for varying values of α. Both plots support previous results that increasing values of stress dissipation decrease the length of the scaling regime and the time of the largest events, and that larger asperity, or damage, regions promote the occurrence of larger events.

In this study we want to investigate the scaling in these systems for different percentages of asperity sites and different asperity configurations, comparing those results with the simple homogeneous system with no asperities.

## 3 Percentage of asperity blocks

A system with different percentage of randomly distributed asperity sites in a two dimensional cellular automaton lattice of linear size $L$=256 with periodic boundary conditions and long range of interaction (R=16) is studied here. In this model we consider a homogeneous failure threshold for the regular sites of $\sigma^F$=2.0, a homogeneous residual stress for the entire lattice of $\sigma^r$=1.0, and a random distribution of noise as $\eta$=[-0.1,+0.1]. The failure threshold for the asperity sites is $\sigma^F_{(asperity)}=\sigma^F+10$. Figure 4 shows the distribution of event sizes for different cases where $\alpha$ is equal to 0.2. This study begins with smaller percentages of asperities and increase the number of randomly distributed stronger sites in the lattice (no asperities in black, 1% in blue, 3% in green and 5% in red). As the percentage of asperities increases, the system produces significantly larger events. However, the relative number of moderate-sized events decreases as the number of asperities in the lattice is increased. By increasing the number of asperities in the lattice some of the moderate events appear to grow into a larger event. This migration from the moderate to large sizes could be consequence of two effects. When an asperity site breaks a greater amount of stress is released into the system and that amount of released stress can cause the failure of more sites and result in larger events, especially in a system with long range stress transfer. In addition, a greater number of randomly distributed asperity sites in the lattice increases the probability of asperity sites triggering. In other words, a system with a higher density of asperities increases the chance of asperity blocks to be in the stress transfer range of another asperity. So, failure of one asperity site can results in a cascade behavior and a greater likelihood for a medium size event to grow and become an extreme event.

## 4    Configuration of spatial heterogeneity

Dominguez et al. (2013) studied the scaling behavior of systems with damage and showed that event distribution depends not only on the total amount of damaged sites in the system but also on the spatial distribution of damage. They noticed that lattices with more homogeneously distributed dead sites suppress large events. Here, we study the effect of different spatial configurations of the asperity blocks in the system. In this model, the lattices have the same size (256 × 256), a constant percentage of asperity sites (1% and 5% for each case), and a constant stress transfer range (R = 16). As a result, any differences in the large event behavior are not caused by the finite size of the lattice. Figure 5 presents two-dimensional lattices of linear size L = 256 and asperity blocks with a linear size of *b* which are randomly distributed throughout the system and two cases of 1% (top) and 5% (bottom) of asperity sites (note that the stress dissipation parameter is constant in all cases, $\alpha = 0.2$). To highlight the distinction between the different configurations of asperities, logarithmic bins are used in the distributions and they are normalized to the total number of events in each case. Figure 6 shows the probability density function of event sizes for the various arrangements of asperity blocks in Figure 5. By changing the linear size of the asperity blocks, we ensure that at least some of the asperity sites which are inside a larger asperity block are in the stress transfer range of others and will certainly interact with each other. On the other hand, we decrease the probability of interaction between two large blocks of asperity sites. Although the number of asperity sites is constant, changing the linear size of the asperity sites has a significant effect on the event distributions. For 1% of asperities, $b=1$ (Figure 5a-i) has the lowest spatial separation among the different size of asperity blocks. But since there are fewer of asperities in the lattice, the probability of interaction between the asperity sites is still very low and big events do not occur even with the failure of an asperity site. For $b=4$, there are 16 asperity sites in a 4 by 4 block that are adjacent to each other. Failing a site in the asperity block can easily trigger the rest of the block and create larger events than for $b=1$ (Figure 6a). The numerical distributions for $b=8$ and $b=16$ confirms that there is triggering of asperity sites inside the block, but because the distance between the blocks is higher than the stress transfer range they cannot affect each other directly. Again, in this case the largest events occur for smaller values of R/$b$.

The biggest event for 5% of asperity sites falls between different configurations of asperity blocks and occurs for the smallest block size (Figure 6b, $b=1$). Because the percentage of asperities is much higher, the shortest spatial separation between the asperities occurs in the case of $b=1$ and there is a higher probability of asperity triggering. In this case, the number of triggered asperity sites increases and a much bigger event occurs. However, increasing the linear size of asperity blocks ($b=4$, 8 and 16) increases the probability of triggering inside a block, but it also increases the spatial separation between the blocks, resulting in a decrease in the probability of triggering between asperity blocks.

## 5  Range of interaction

To further investigate the effect of block size and asperity triggering on event size for larger numbers of asperities, we focus on the stress transfer range. Here, 5% of asperity sites are randomly distributed in the system in the blocks with four different linear sizes ($b$=1, 4, 8 and 16, Figure 5a-ii, b-ii, c-ii and d-ii) and the system is run for three different stress transfer ranges, R=16, R=8 and R=1. Figure 7 shows the comparison between the event distributions for different stress transfer ranges and different asperity block sizes. By reducing the stress transfer range, the likelihood that an asperity site will be affected by a neighboring asperity site decreases. This results in a lower probability of asperity triggering in the model. The absence of triggering asperities makes the system unable to produce big events. Figure 7 confirms that, for lower percentages of asperities, the size of the biggest event occurs in systems with long range interactions (R=16), regardless of the asperity block size, and that event is much larger than the size of the biggest event in system with short stress transfer range. In addition, for short range stress transfer (R=1), the size of the largest events does not change with increasing asperity block size. For longer range stress transfer, the size of the asperity blocks affects the size of the biggest event.

## 6  Stress dissipation

We also compare the effect of stress dissipation parameters on different asperity configurations in a system with long stress transfer range, R=16. As discussed earlier, upon failure, some fraction of a site's stress, given by $\alpha\,[\sigma^F-(\sigma^r\pm\eta)]$, is dissipated from the system ($0<\alpha\leq1$). In general, lower stress dissipation models produce more larger events and higher stress dissipation suppresses large events. This also should be true in the system with asperity sites. In higher dissipation models, less stress is transferred to neighboring sites, even when asperities fail. As a result, there is lower probability of asperity triggering in the model. This implies smaller events and more plate update steps to fail all of the asperities. In Figure 8 the event distribution for four different stress dissipation parameters and 1% of asperity sites in the lattice is compared for three different configuration of asperity blocks ($b=1$, $b=4$ and $b=16$). This figure shows the results for four stress dissipation parameters, $α=0.2$, 0.3, 0.4 and 0.5, in each 20%, 30%, 40% and 50% of the failed site stress dissipates at the time of failure and the remaining stress is distributed to its $z=1088$ neighbors. For all three asperity configurations (Figure 8 a, b and c), lower values of stress dissipation leads to a higher number of big events. On the other hand, in systems with a high amount of stress dissipation, the released stress from a small asperity block is not enough to offset the greater dissipation and cannot trigger any of the neighboring asperity blocks. As a result, the change in the asperity block size from $b=1$ to $b=4$ (Figure 8a and 8b) does not significantly affect the distributions of events, especially for higher stress dissipation. However, the released stress from bigger asperity blocks of $b=16$ is high enough to trigger another asperity block. We again observe larger events in the tail of the distributions, even for higher stress dissipation models (Figure 8c).

## 7  Conclusions

Here we investigated a variation of the OFC model in which localized stress accumulators were added to the system by converting a small percentage of the lattice site into stronger sites. We studied different spatial configurations of the stronger asperity sites and observed the effect of asperity patterns on the distribution of event sizes in the systems for a selected stress transfer range. In particular, we observed an increasing number of larger events associated with the total number of asperities in the lattice (Figure 4). We also found that imposing a fixed number of asperity sites with different spatial distributions strongly affects the capability of the fault system to generate extreme events, but that the total percentage of asperities is important as well. The event distributions shown in Figure 6 confirm the sensitivity of the system to different configurations of inhomogeneities. In addition, we studied the role of the interaction range on triggering asperities and observed that the probability of asperity triggering and the occurrence of larger events is much higher in the in those systems with a longer stress transfer range (Figure 7). However, in models with higher dissipation it is the asperities are less sensitive to other failures and therefore the probability of asperity triggering decreases in high dissipation models. This implies that higher dissipation results in a lower probability of smaller events regardless of the number and spatial distribution of asperities.

## Acknowledgements

This research was funded by the NSERC and Aon Benfield/ICLR IRC in Earthquake Hazard Assessment, and an NSERC Discovery Grant (JK and KFT).

**References**

Alava, M. J., Nukala, P., and Zapperi, S. (2006), Statistical models of fracture, Adv. Phys. 55, 349.

Bach, M., Wissel F. and Dressel, B. (2008), Olami-Feder-Christensen model with quenched disorder. Phys. Rev. E77, 067101.

Burridge, R., Knopoff, L. (1967), Model and theoretical seismicity, Bull. Seismol. Soc. Am. 57. 341–371.

Carlson, J. M. and Langer, J. S. (1989), Mechanical model of an earthquake fault. Phys. Rev. Lett. 62, 2632; Phys. Rev. A 40, 6470.

Ceva, H. (1995), Inhuence of defects in a coupled map lattice modelling earthquakes. Phys. Rev, E52, 154.

Dahmen, K., Ertas, D., and Ben-Zion, R. (1998), Gutenberg-Richter and Characteristic Earthquake Behavior in Simple Mean-field Models of Heterogeneous Faults, Phys. Rev. E 58, 1494–1501.

Dominguez, R., Tiampo, K.F., Serino, C.A. and Klein W. (2013), Scaling of earthquake models with inhomogeneous stress dissipation. Phys. Rev. E 87, 022809.

Gutenberg, B. and Richter, C.F. (1956), Magnitude and energy of earthquakes. Annali di Geofisica, Vol. 9, n. 1.

Herz, A.V.M., and Hopfield, J.J. (1995), Earthquake Cycles and Neural Reverberations, Collective Oscillations in Systems with Pulse-Coupled Threshold Elements. Phys. Rev. Lett., 75, 1222-1225.

Jagla, E.A. (2010), Realistic spatial and temporal earthquake distributions in a modified Olami-Feder-Christensen model. Phys. Rev, E81, 046117.

Janosi, I.M. and Kertesz, J. (1994), Self-organized criticality with and without conservation. Physica A200, 0378.

Kanamori, H. (1981), The nature of seismicity patterns before large earthquakes, in Earthquake Prediction, Maurice Ewing Series, IV, 1–19, AGU, Washington D.C..

Klein, W., Gould, H., Gulbahce, N., Rundle, J. B. and Tiampo, K. (2007), Structure of fluctuations near mean-field critical points and spinodals and its implication for physical processes. Phys. Rev. E 75, 031114.

Lyakhovsky, V. and Ben-Zion, Y. (2009), Evolving geometrical and material properties of fault zones in a damage rheology model, Geochem. Geophys. Geosyst., 10, Q11011.

Main, I. (1996), Statistical physics, seismogenesis, and seismic hazard. Rev. Geophys. 34, 433–462.

Mousseau, N. (1996), Synchronization by Disorder in Coupled Systems. Phys. Rev. Lett. 77, 968.

Nakanishi, H. (1990), Cellular-automaton model of earthquakes with deterministic dynamics. Phys. Rev. A 41, 7086.

Olami, Z., Feder, HJS., Christensen, K. (1992), Self-organized criticality in a continuous, nonconservative cellular automaton modelling earthquakes. Phys Rev Lett 68(8):1244–1247.

Ogata, Y. (1983), Estimation of the parameters in the modified Omori formula for aftershock frequencies by the maximum likelihood procedure, Journal of Physics of the Earth, Vol.31, 115-124.

Otsuka, M. (1972), Phys Earth Planet Inter. 6-311.

Ramos, O., Altshuler, E. and Maloy K.J. (2006), Quasiperiodic Events in an Earthquake Model. Phys. Rev. Lett. 96, 098501.

Rundle, J.B., Turcotte, D.L., Shcherbakov, R., Klein, W., Sammis, C.,(2003), Statistical physics approach to understanding the multiscale dynamics of earthquake fault systems. Review of Geophysics, 41, 1019.

Rundle, J. B., Klein, W., Tiampo, K. and Gross S., (2000), Linear pattern dynamics in nonlinear threshold systems, Phys. Rev. E, 61 (3), 2418–2431.

Rundle, JB, Klein W. and Gross S. (1999), Physical basis for statistical patterns in complex earthquake populations: Models, predictions, and tests, PAGEOPH, 155, 575-607.

Rundle, J. B. and Brown, S. R. (1991), Origin of Rate Dependence in Frictional Sliding, J. Stat. Phys. 65, 403.

Rundle, J. B. (1988), A physical model for earthquakes, J. Geophys. Res. 93-6237.

Rundle, J. B. and Jackson, D. D. (1997), Numerical simulation of earthquake sequences, Bull. Seismol. Soc. Am. 67.

Serino, C. A., Tiampo, K. F. and Klein W. (2011), New Approach to Gutenberg-Richter Scaling, Phys. Rev. Lett. 106, 108501.

Scholz, C.H. (2002), The mechanics of earthquakes and faulting, Cambridge University Press, p. 471.

Schorlemmer, D., and Gerstenberger, M. C. (2007), RELM testing center, Seismol. Res. Lett. 78, 1, 30-36.

Tiampo, K. F., Rundle, J. B., Klein, W., Holliday, J., S´aMartins, J. S. and Ferguson, C. D. (2007), Ergodicity in natural earthquake fault networks. Phys. Rev. E 75, 066107.

Tiampo, K.F., Rundle, J.B., McGinnis, S., Gross, S., Klein, W. (2002), Mean-field threshold systems and phase dynamics: an application to earthquake fault systems. Eur. Phys. Lett. 60, 481–487.

Torvund F., and Froyland, J. (1995), Strong ordering by non-uniformity of thresholds in a coupled map lattice. Physica Scripta 52, 624.

Turcotte, D.L., Newman, W.I., Shcherbakov, R. (2003), Micro and macroscopic models of rock fracture. Geophysical Journal International 152, 718–728.

Turcotte, D. L. (1997), Fractals and chaos in geology and geophysics, 2nd edn. Cambridge, UK: Cambridge University Press.

Utsu, T., Ogata, Y. and Matsu'ura, R. S. (1995), The centenary of the Omori formula for a decay law of aftershock activity, Journal of Physics of the Earth, Vol.43, pp.1-33.

Vere-Jones, D. (2006), The development of statistical seismology, A personal experience. Tectonophysics 413, 5–12.

Vere-Jones, D. (1995), Forecasting earthquakes and earthquake risk. International Journal of Forecasting 11, 503–538.

Zechar, J.D., Jordan, T.H. (2010), Simple smoothed seismicity earthquake forecasts for Italy. Annals of Geophysics 53.

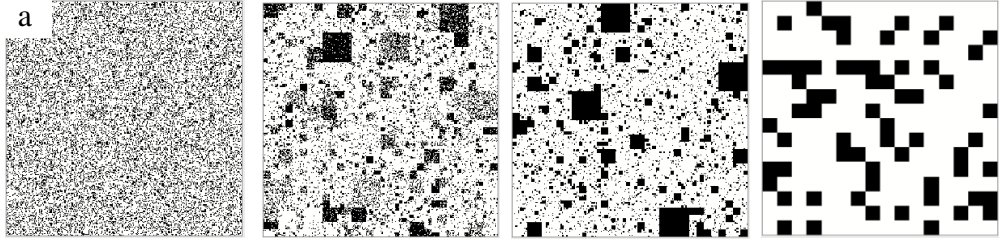

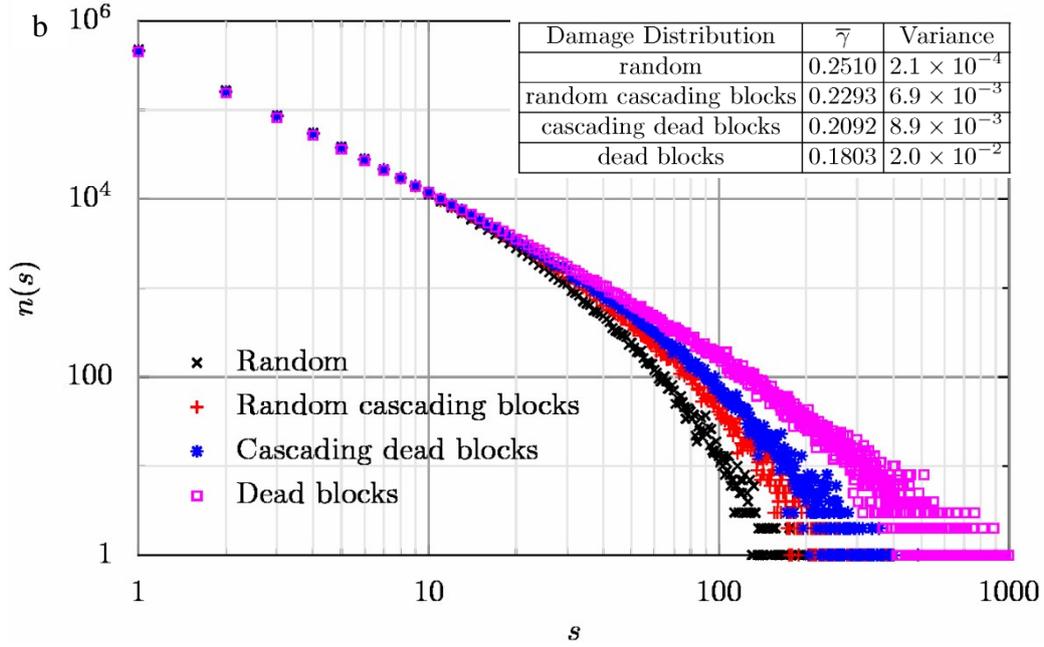

Figure 1. a) Various configurations of 25% dead sites (in black) for a lattice with α = 0 and a linear size L = 256. From left to right, lattices contain dead sites distributed randomly (random), blocks of various sizes, where each block has varying values of randomly distributed dead sites (random cascading blocks), completely dead blocks of various sizes (cascading dead blocks), and dead blocks of a uniform size (dead blocks); b) Numerical distribution of avalanche events of size s for the various spatial distributions of dead sites shown in (a) with stress transfer range R = 16. $\bar{\gamma}$ is the average dissipation for each lattice.

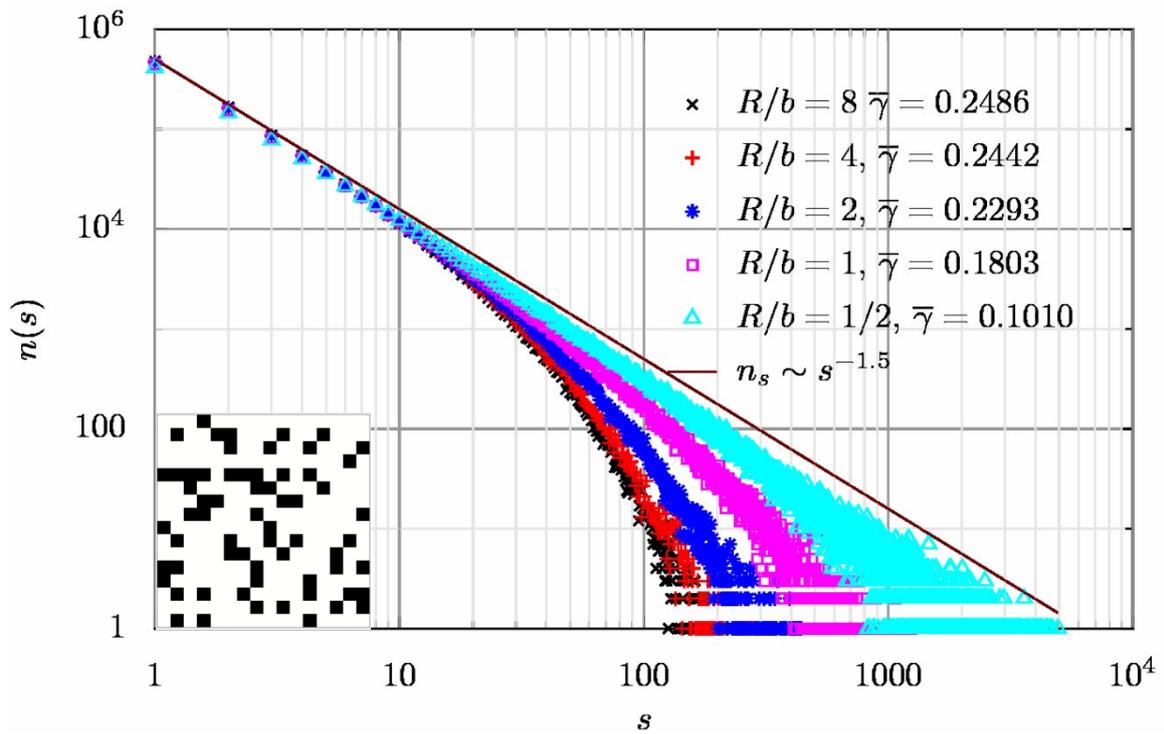

Figure 2. Numerical distribution of avalanche events of size s for blocks of dead sites of linear size b. Systems are characterized by the dimensionless parameter R/b. (the inset corresponds to R/b = 1.) The size of the systems shown here is L = 256, α = 0, and the stress transfer range is R = 16.

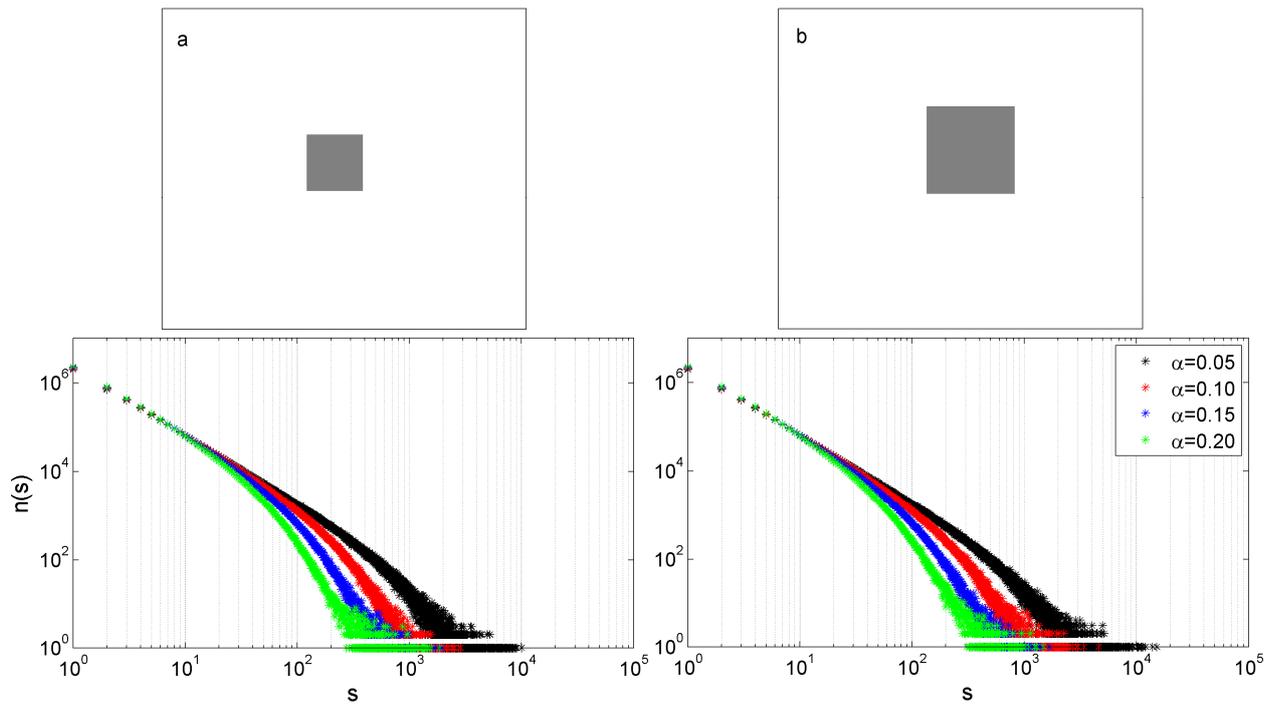

Figure 3. a) Schematic depicting the size and location of a smaller set (5% of the total lattice sites) of grouped asperities (top). Numerical distribution of event sizes for varying values of stress dissipation, α, for the layout shown above. b) Schematic depicting the size and location of the larger set (10% of the total lattice sites) of grouped asperities (top). Numerical distribution of event sizes for varying values of stress dissipation, α, for the layout shown above.

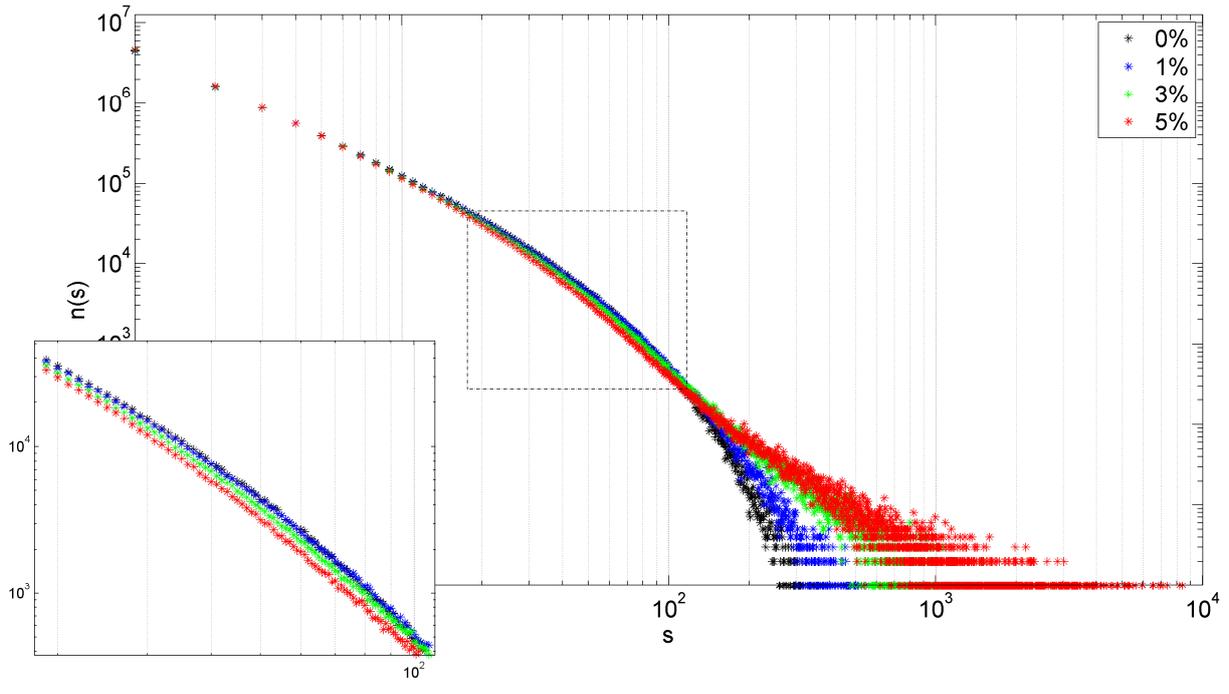

Figure 4. Numerical distribution of events of size "s" for various amounts of randomly distributed asperity sites, $α=0.2$.

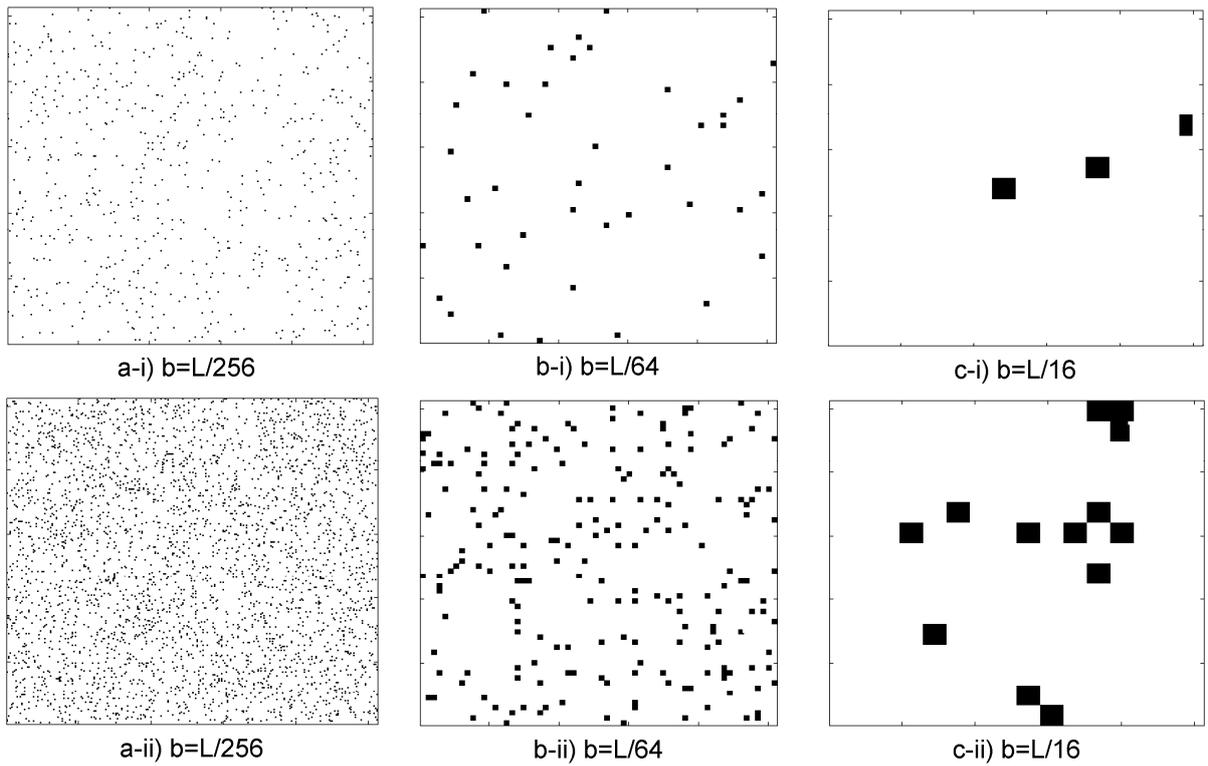

Figure 5. Four different configurations of 1 percent asperity sites (upper row) and 5 percent asperity sites (lower row) (in black) for lattices with linear size $L$=256. Each lattice has asperity blocks of a single size b. For each **a**) $b=L/256$, **b**) $b=L/64$, **c**) $b=L/32$, and **d**) $b=L/16$.

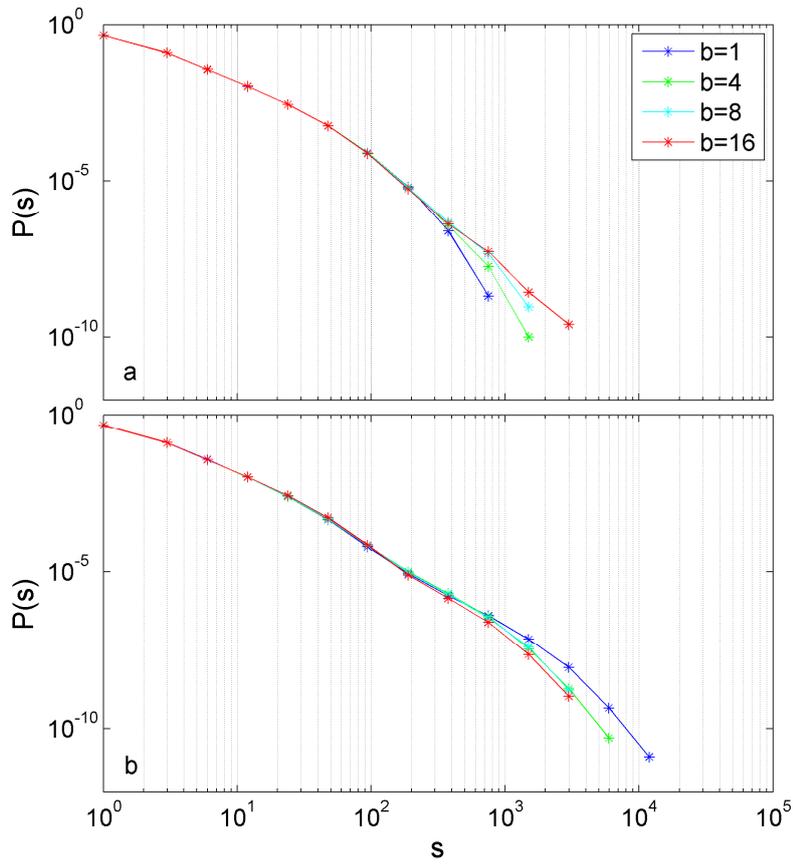

Figure 6. Normalized probability density function of events of size "s" for various sizes of asperity blocks which are randomly distributed in the system for R=16 and **a**) 1% and **b**) 5% of asperity sites.

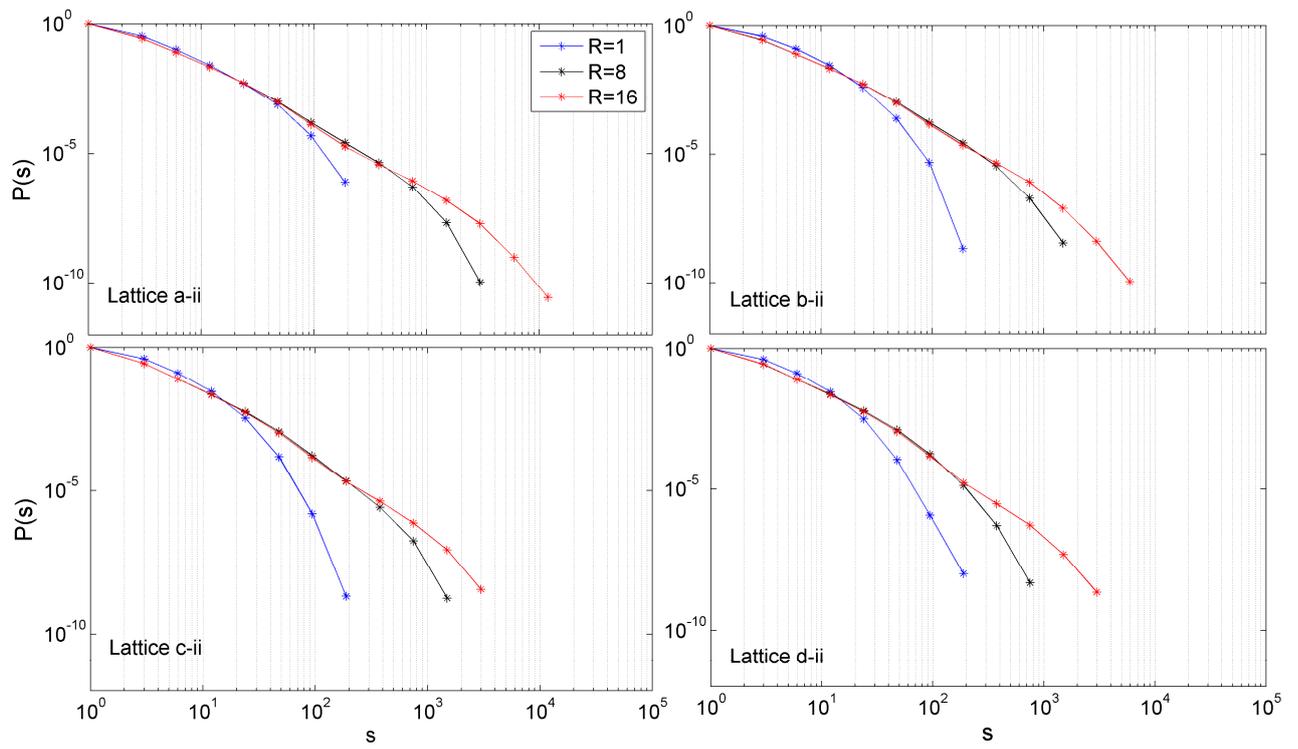

Figure 7. Normalized probability density function of events of size s for various stress transfer range in a system with 5% of asperity sites for four different linear sizes of asperity blocks shown in Figure 2 and **a**) $b=1$, **b**) $b=4$, **c**) $b=8$ and **d**) $b=16$.

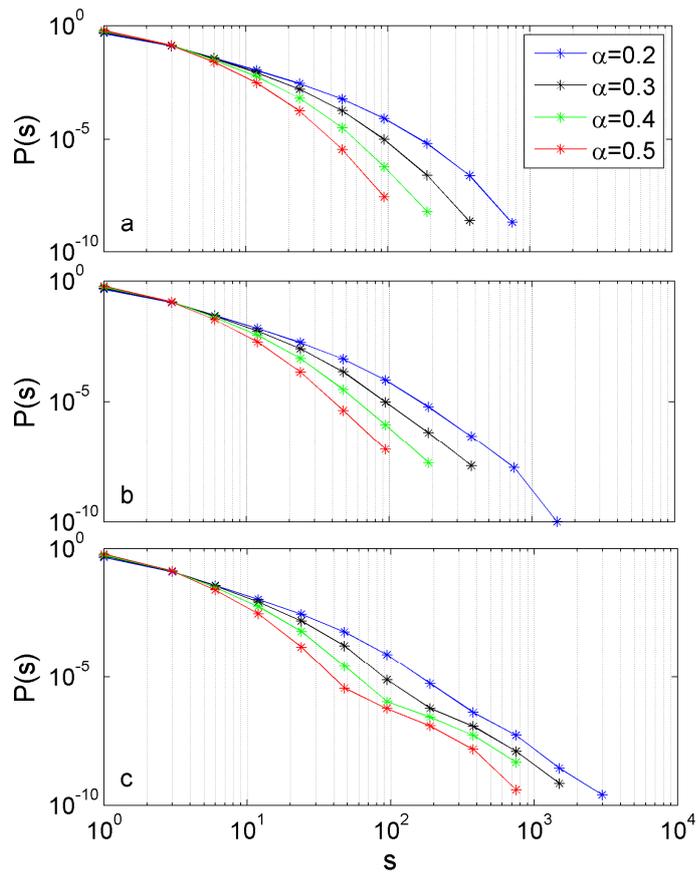

Figure 8. Normalized probability density function of events of size s for various amount of stress dissipation parameter and for 1% of three different sizes of asperity blocks randomly distributed in the system for **a)** $b=1$, **b)** $b=4$ and **c)** $b=16$.